\documentclass[11pt,a4paper]{amsart}
\usepackage {graphicx,epsfig,float}
\usepackage{amsmath}
\usepackage{graphicx}
\usepackage{amsfonts}
\newcommand{\R}{{\Bbb R}}

\newcommand{\ord}[1]{\ensuremath{ {\text \em o} \!\left(#1 \right)} }

\newcommand{\D}[2]{ \ensuremath{ \frac{d #1 }{d #2 } }}

\begin{document}
\title[Dynamical robustness of biological
networks]{Dynamical robustness of biological networks with
hierarchical distribution of time scales}

\author{A.N.Gorban}\thanks{ANG: University of Leicester, Leicester,
LE1 7RH, UK, e-mail: ag153@le.ac.uk}

\author{O.Radulescu}\thanks{OR: IRMAR, UMR CNRS 6625,  Universit\'e de Rennes 1,
France, \\ e-mail: ovidiu.radulescu@math.univ-rennes1.fr}

\date{}

\begin{abstract}
We propose the concepts of distributed robustness and
$r$-robustness, well adapted to functional genetics. Then we discuss
the robustness of  the relaxation time using a chemical reaction
description of genetic and signalling networks. First, we obtain the
following result for linear networks: for large multiscale systems
with hierarchical distribution of time scales the variance of the
inverse relaxation time (as well as the variance of the stationary
rate) is much lower than the variance of the separate constants.
Moreover, it can tend to $0$ faster than $1/n$, where $n$ is the
number of reactions. We argue that similar phenomena are valid in
the nonlinear case as well. As a numerical illustration we use a
model of signalling network that can be applied to important
transcription factors such as NF$\kappa$B.

\end{abstract}

\maketitle

\section{Introduction}

Robustness, defined as stability against external perturbations and
internal variability, represents a common feature of living systems.
 The fittest organisms are those organisms that resist to
diseases, to imperfections or damages of regulatory mechanisms, and
that can function reliably in various conditions. There are many
theories that describe, quantify and explain robustness.
Waddington's canalization, rigorously formalized by Thom describes
robustness like structural stability of attractors under
perturbations \cite{Thom1,wad57}. Many useful ideas on robustness
have been imported from the theory of control of dynamical systems
and of automata \cite{ fates, chaves}. The new field of systems
biology places robustness in a central position among the living
systems organizing principles, identifying redundancy, modularity,
and negative feed-back as sources of robustness
\cite{kitano,kitano04a,kitano04b}.

In this paper, we provide some justification to a different, less
understood source of robustness.

Early insights into this problem can be found in the von Neumann's
\cite{vonneumann} discussion of robust coupling schemes of
automata. von Neumann noticed the intrinsic relation between
randomness and robustness. Quoting him ``without randomness,
situations may arise where errors tend to be amplified instead of
cancelled out; for example it is possible, that the machine
remembers its mistakes, and thereafter perpetuates them". To cope
with this, von Neumann introduces multiplexing and random
perturbations in the design of robust automata.

Related to this is Wagner's \cite{wagner,wagner05} concept of
distributed robustness which ``emerges from the distributed nature
of many biological systems, where many (and different) parts
contribute to system functions". To a certain extent, distributed
robustness and control are antithetical. In a robust system, any
localized perturbation should have only small effects. Robust
properties should not depend on only one, but on many components
and parameters of the system. A weaker version of distributed
robustness is the $r$-robustness, when $r$ or less changes have
small effect on the functioning of the system \cite{Deutscher}.

Molecular biology offers numerous examples of distributed
robustness and of $r$-robustness. Single knock-outs of
developmental genes in the fruit fly have localized effects and do
not lead to instabilities \cite{bahram02}. Complex diseases are
the result of deregulation of many genetic pathways
\cite{coronary}. Transcriptional control of metazoa  is based on
promoter and enhancer regulating DNA regions that collect
influences from many proteins \cite{ptashne_book}. Networks of
regulating micro-RNA could be key players in canalizing genetic
development programs \cite{hornstein}. Interestingly, computer
models of gene regulation networks \cite{vonDassow00} have
distributed robustness with respect to variations of their
parameters. Flux balance analysis in-silico studies of the effects
of multiple knock-outs in yeast {\em S.cerevisiae} identified sets
of up to eight interacting genes and showed that yeast metabolism
is less robust to multiple attacks than to single attacks
\cite{Deutscher}.

Let us formulate the problem mathematically. A property $M$ of the
biological system is a function of several parameters of the
system, $M=f(K_1,K_2,\ldots,K_n)$. Let us suppose that the
parameters $(K_1,K_2,\ldots,K_n)$ are independent random
variables. There are various  causes of variability: mutations,
across individuals variability, changes of the functional context,
etc. For definitions of robustness we can start from inequality:
$Var(M) \ll Var(K_i),\,  1 \leq i \leq n$.

In order to avoid the problem of units and  supposing that
$M,K_i>0$, we can use logarithmic scale:

{\bf Definition 1}: $M$ is {\em robust with respect to distributed
variations} if the variance of $M$ is much smaller than the
variance of any of the parameters:

\begin{equation}
Var(\log M) \ll Var(\log K_i),\,  1 \leq i \leq n. \label{eq1bis}
\end{equation}

We can also say that say that $M$ ``concentrates" on its central
value $M_c$.

Let us consider $r$-index subsets $I_r = \{ i_1,i_2,\ldots,i_r \}
\subset \{ 1,2,\ldots,n \}$ for given $r$. Let
$K_i^0,i=1,\ldots,n$ be the central values of the parameters. For
given $I_r$, the perturbed values $K_i$ are obtained by
multiplying $r$ selected central values by independent random
scales $s_i > 0, i=1,\ldots,r$, $K_i = K_i^0 s_i, i \in I_r$.

{\bf Definition 2}: $M$ is {\em robust with respect to $r$
variations} or {\em $r$-robust} if for any $I_r$:
\begin{equation}
Var(\log M) \ll Var(\log s_i),\,  i \in I_r. \label{eq3}
\end{equation}

{\em $r$-robustness} holds if  \eqref{eq3} is valid for any
deterministic choice of $r$ targets.  If the  target set $I_r$ is
randomly chosen we shall  speak of {\em  weak $r$-robustness}. We
call {\em robustness index} the maximal value of $r$ such that the
system is $r-$robust.

The above definitions are inspired from biological ideas. Our first
definition corresponds to Wagner's distributed robustness
\cite{wagner05}. It expresses the fact that $M$ is not sensitive to
random variations of the parameters. $r$-robustness has been defined
in \cite{Deutscher} as resistance with respect to multiple
mutations. $r$-robustness can also be interpreted as functional
redundance \footnote{This is different from the structural
redundance of Wagner \cite{wagner05}, meaning that many genes code
for the same protein.} meaning that the property $M$ is collectively
controlled by more than $r$ parameters, and can not be considerably
influenced by changing a number of parameters less than or equal to
$r$. One should also notice the introduction of a new concept. Even
if there are $r$ critical targets (for instance genes whose
mutations lead to large effects) the probability of hitting these
$r$ targets randomly, could be small. We have introduced the weak
$r$-robustness to describe this situation.

Robustness with respect to distributed variations defined by can
be a consequence of the Gromov-Talagrand concentration of measure
in high dimensional metric-measure spaces \cite{Talagrand,Gromov}.
In Gromov's theory the concentration has a geometrical
significance: objects in very high dimension look very small when
they are observed via the values of real functions
(1-Lipschitzian).  This represents  an important generalization of
the law of large numbers and  has many applications in
mathematics.

In this paper we choose a signaling module example as an
illustration of the concept of distributed robustness. The robust
property that we study here is the relaxation time of a biological
molecular system modeled as a network of chemical reactions.
Relaxation time is an important issue in chemical kinetics, but
there exists biological specifics.  A biological system is a
hierarchically structured open system. Any biological model is
necessarily a submodel of a bigger one. After a change of the
external conditions, a cascade of relaxations takes place and the
spatial extension of a minimal model describing this cascade
depends on time. Timescales are important in signalling between
cells and between different parts of an organism. It is therefore
important to know how the relaxation time depends on the size and
the topology of a network and how robust is this time against
variations of the kinetic constants.

The structure of this paper is the following. First, we extend the
classical results on limiting steps of stationary states of
one-route cyclic linear networks onto dynamic of relaxation of any
linear network. This allows us to relate the relaxation time of a
linear network with hierarchical distribution of time scales to
low order statistics of the network constants and to prove the
distributed robustness of this relaxation time. Last, using a
model of the NF$\kappa$B signaling module as an example, we show
that similar results apply to nonlinear networks. For this
nonlinear network, the robustness of another characteristic time,
the period of its oscillations is studied as well.

\section{Limitation of relaxation in linear reaction networks}

First we consider a linear network of chemical reactions. In a
linear network, all the reactions are of the type $A_i \rightarrow
A_j$ and the reaction rates are proportional to the reagents $A_i$
concentration.

The dynamics of the network is described by:

\begin{equation} \label{dynamics}
\dot{c}=Kc
\end{equation}

where $K$ is the matrix of kinetical parameters.

We call a linear network  {\em weakly ergodic },  if for any initial
state $c(0)$ there exists a limit state $\lim_{t\rightarrow \infty }
\exp(Kt) \, c(0)$ and the set of all these limit states for all
initial conditions is a one-dimensional subspace (for all $c$,
$\lim_{t\rightarrow \infty } \exp(Kt) \, c = \lambda(c) c^*$,
$\lambda(c) \in \R$.

The ergodicity of the network follows from its topological
properties.

A non-empty set $V$ of graph vertexes forms a {\it sink}, if there
are no oriented edges from $A_i \in V$ to any $A_j \notin V$. For
example, in the reaction graph $A_1\leftarrow A_2 \rightarrow A_3$
the one-vertex sets $\{A_1\}$ and $\{A_3\}$ are sinks. A sink is
minimal if it does not contain a strictly smaller sink. In the
previous example, $\{A_1\}$, $\{A_3\}$  are minimal sinks. Minimal
sinks are also called ergodic components.

A linear conservation law is a linear function defined on the
concentrations $b(c)= \sum_{i=1}^n b_i c_i$, whose value is
preserved by the dynamics \eqref{dynamics}. The set of all the
conservation laws forms the left kernel of the matrix $K$.

From ergodic Markov chain theory it follows that the following
properties are equivalent:

i) the network is weakly ergodic.

ii) for each two vertices  $A_i, \: A_j \: (i \neq j)$ we can find
such a vertex $A_k$ that oriented paths exist from $A_i$ to $A_k$
and from $A_j$ to $A_k$. One of these paths can be degenerated: it
might be $i=k$ or $j=k$.

iii) the network has only one minimal sink (one ergodic component).

iv) there is an unique linear conservation law, namely $b(c)=
\sum_{i=1}^n c_i$.

Hence, the maximal number of independent linear conservation laws is
equal to the maximal number of disjoint ergodic components.

Now, let us suppose that the kinetic parameters are well separated
and let us sort them in decreasing order : $k_{(1)} \gg k_{(2)}
\gg \ldots \gg k_{(n)}$. Let us also suppose that the network has
only one ergodic component (when there are several ergodic
components, each one has its longest relaxation time that can be
found independently). We say that $k_{(r)}, \, 1 \leq r \leq q$ is
the
  {\em ergodicity boundary}
  if the network of  reactions with parameters $k_1,k_2, \ldots , k_r$ is weakly ergodic,
but the network with parameters $k_1,k_2, \ldots , k_{r-1}$ it is
not. In other words, when eliminating reactions in decreasing order
of their characteristic times, starting with the slowest one, the
ergodicity boundary is the constant of the first reaction whose
elimination breaks the ergodicity of the reaction graph.

Let $\lambda_i$ be the eigenvalues of the matrix $K$ (these satisfy
$Re \lambda_i \leq 0$; furthermore if $Re \lambda_i  = 0$ then
necessarily $\lambda_i = 0$). Relaxation to equilibrium of the
network is multi-exponential, but the longest relaxation time is
given by :

\begin{equation}
\tau = 1 / \min \{ - Re \lambda_i |  \lambda_i \neq 0 \}
\end{equation}

An estimate of the longest relaxation time can be obtained by
applying the perturbation theory for linear operators to the
degenerated case of the zero eigenvalue of the matrix $K$. We have
$K=K_{<r}(k_1,k_2,\ldots,k_{r-1})+k_rQ$, where $K_{<r}$ is
obtained from $K$ by letting $k_r=k_{r+1}=\ldots k_n=0$, $Q$ is a
constant matrix up to terms that are negligible relative to $k_r$.
From Lemma, the zero eigenvalue is twice degenerated in $K_{<r}$
and only once degenerated in $K$. One gets the following estimate:

\begin{equation}
\overline{a}\frac{1}{k_{(r)}} \geq \tau \geq
\underline{a}\frac{1}{k_{(r)}}, \label{eq10}
\end{equation}
where $\overline{a}, \underline{a} >0$ are some positive functions
of $k_1,k_2, \ldots , k_{r-1}$ (and of the reaction graph topology).

Two simplest examples give us the structure of the perturbation
theory terms for $\min_{\lambda \neq 0} \{- Re \lambda \}$.
\begin{figure}[H]
\begin{centering}
a)\includegraphics[width=26mm, height=23mm]{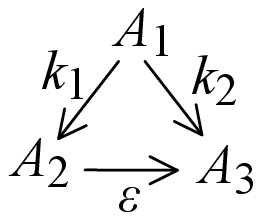}
b)\includegraphics[width=26mm, height=23mm]{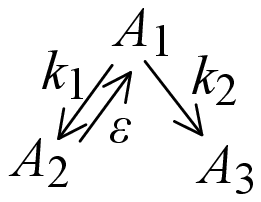}
\caption {\label{Fig:Triangl} Two basic examples of ergodicity
boundary reaction: (a) Connection between ergodic components; (b)
Connection from one ergodic component to element that is connected
to the both ergodic components by oriented paths. In both cases,
for $\varepsilon=0$, the ergodic components are $\{A_2\}$ and
$\{A_3\}$.}
\end{centering}
\end{figure}

\begin{enumerate}
\item{Connection between ergodic components. For the reaction mechanism
Fig.~\ref{Fig:Triangl}a $\min_{\lambda \neq 0} \{- Re \lambda
\}=\varepsilon$, if $\varepsilon < k_1+k_2$.}
\item{Connection from one ergodic component to element that is connected to the both ergodic
components by oriented paths. For the reaction mechanism
Fig.~\ref{Fig:Triangl}b $\min_{\lambda \neq 0} \{- Re \lambda
\}=\varepsilon k_2/(k_1+k_2) + \ord{ \varepsilon}$, if
$\varepsilon < k_1+k_2$. For well separated parameters there
exists a ``trigger" alternative: if $k_1 \ll k_2$ then
$\min_{\lambda \neq 0} \{- Re \lambda \}\approx \varepsilon$; if,
inverse, $k_1 \gg k_2$ then $\min_{\lambda \neq 0} \{- Re \lambda
\}= \ord{ \varepsilon}$.}
\end{enumerate}

More generally, let us suppose that the ergodicity boundary
 $k_{(r)}$ corresponds to a reaction that connects two disjoint
 ergodic components of the reaction graph
 $k_{(1)},\ldots,k_{(r-1)} $. Then we have:

\begin{equation}
\tau = \frac{1}{k_{(r)}} +  \ord{ \frac{1}{k_{(r)}} } \label{eq11}
\end{equation}

If the ergodicity boundary  $k_{(r)}$ corresponds to a reaction
that connects one ergodic component to element that is connected
to the both ergodic components by nonempty oriented paths (as in
Fig.~\ref{Fig:Triangl}b), then we have the trigger alternative
again, $1/ \tau$ is either $k_{(r)}$, or $\ord{k_{(r)}}$.

Thus, the well known concept of stationary reaction rates {\it
limitation} by ``narrow places" or ``limiting steps" (slowest
reaction) should be complemented by the {\it ergodicity boundary}
limitation of relaxation time. It should be stressed that the
relaxation process is limited  not by the classical limiting steps
(narrow places), but by  reactions that may be absolutely different.
The simplest example of this kind is a catalytic cycle: the
stationary rate is limited by the slowest reaction (the smallest
constant), but the relaxation time is limited by the reaction
constant with the second lowest value (in order to break the
ergodicity of a cycle two reactions must be eliminated).

\section{Robustness of relaxation time in  linear systems}

In general, for large multiscale systems we observe concentration
effects: the log-variance of the relaxation time  is much lower
than the log-variance of the separate constants. For linear
networks, this follows from well known properties of the order
statistics \cite{lehmann}. For instance, if $k_i$ are independent,
log-uniform random variables, we have $Var[\log(k_{(r)})] \sim
1/n^2$. Here we meet a ``simplex--type" concentration
(\cite{Gromov}, pp. 234--236) and the log-variance of the
relaxation time can tend to $0$ faster than $1/n$, where $n$ is
the number of reactions.

For log-uniform parameters, $k_{(r)}$ has a  log-beta distribution
$\log(k_{(r)}) \sim Beta(r,n+1-r)$.

We can obtain design principles for robust networks. Suppose we have
to construct a linear chemical reaction network. How to increase
robustness of the largest relaxation times for this network?  To be
more realistic let us take into account two types of network
perturbation:
\begin{enumerate}
\item{random noise in constants;}
\item{elimination of a link or of a node in reaction network.}
\end{enumerate}

Long routes are more robust for the perturbations of the first kind.
So, the first receipt is simple: let us create long cycles! But long
cycles are destroyed by link or node elimination. So, the second
receipt is also simple: let us create a system with many alternative
routes!

Finally the resources  are expensive, and we should create  a
network of minimal size.

Hence, we come to a new combinatorial problem. How to create a
minimal network that satisfies the following restrictions
\begin{enumerate}
\item{the length of each route is $>L$;}
\item{after destruction of $D_{\rm l}$ links and $D_{\rm n}$ nodes
there remains at least one route in the network.}
\end{enumerate}

In order to obtain the minimal network that fulfills the above
constraints, we should include bridges between cycles, but the
density of these bridges should be sufficiently low in order not to
affect significantly the length of the cycles.

Additional restrictions could be involved. For example, we can
discuss not all the routes, but productive routes only (that produce
something useful).

For acyclic networks, we obtain similar receipts: long chains should
be combined with bridges. A compromise between the chain length and
number of bridges is needed.

We can also mention the role of degradation reactions (reaction with
no products). Concentration phenomena are more accentuated when the
number of degradation processes with different relaxation times is
larger. Thus, one can increase robustness by increasing the spread
of the lifetimes of various species.

The detailed discussion of this problem will be published
separately.

\section{Robustness of characteristic times in nonlinear systems: an example}

\subsection*{The model}
Our example is one of the most documented transcriptional regulation
systems in eukaryote organisms:  the signalling module of
NF$\kappa$B. The response of this factor to a signal has been
modeled by several authors
\cite{hoffmann02,lipniacki04,nelson04,ihekwaba04}.

The transcription factor NF$\kappa$B is a  protein (actually a
heterodimer made of two smaller molecules p50 and p65) that
regulates the activity of more than one hundred genes and other
transcription factors that are involved in the immune and stress
response, apoptosis, etc. NF$\kappa$B is thus the principal mediator
of the response to cellular agression and is activated by more than
150 different stimuli : bacteria, viral and bacterial products,
mitogen agents, stress factors (radiations, ischemia, hypoxia,
hepatic regeneration, drugs among which some anticancer drugs).
 NF$\kappa$B has complex regulation, including inhibitor degradation
 and production, translocation between nucleus and cytoplasm,
negative and positive feed-back. Under normal conditions,
NF$\kappa$B is trapped in the cytoplasm where it forms a molecular
complex with its inhibitor I$\kappa$B. Under this form, NF$\kappa$B
can not perform its regulatory function, the complex can not
penetrate the nucleus. A signal, that can be modeled by a kinase
(IKK) frees NF$\kappa$B by degrading its inhibitor. Free NF$\kappa$B
enters the nucleus and regulates the transcription of many genes,
among which the gene of its inhibitor I$\kappa$B and the gene of a
protein $A20$ that inactivates the kinase.

\begin{figure}
\centerline{
\includegraphics[height=14cm]{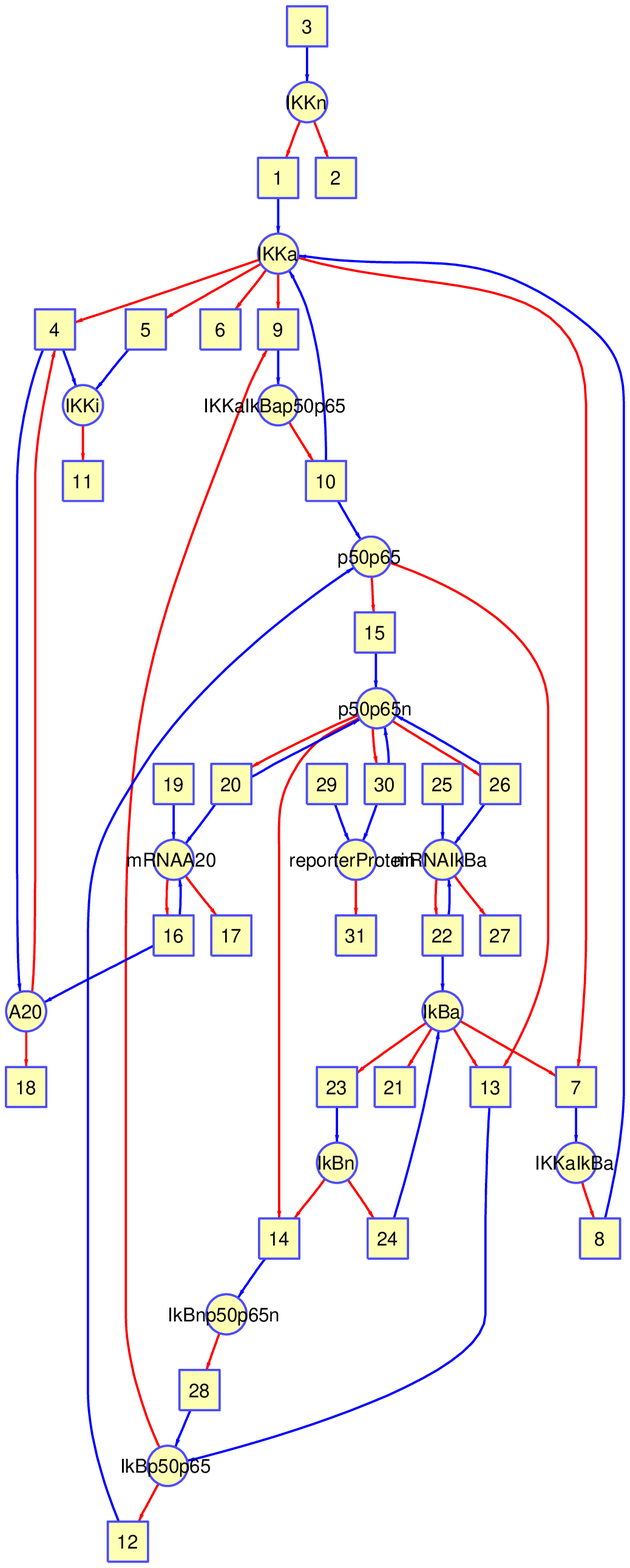}}

\caption{ Model of NF$\kappa$B signalling. The non-linear reaction
mechanism is represented as bipartite graph. The red edges indicate
consumption, while the blue edges indicate production of molecules.
There are $15$ chemical species and $31$ reactions.
 } \label{lipni-model}
\end{figure}

Explaining context-dependent behavior of signal transduction
pathways represents a great challenge in physiology. A rather
popular explanation is pathway crosstalk. Another appealing
explanation is that the signal could carry more information than
it is expected due to its synchronous or asynchronous periodic
nature \cite{nelson04}. These theories should be assessed
experimentally and it is not our purpose to discuss their
validity. Here we would like to study the robustness of the
characteristic times of a non-linear molecular system. In
particular, the double negative feed-back (via $I\kappa B$ and
$A20$) is responsible for oscillations of NF$\kappa$B activity
under persistent stimulation
\cite{hoffmann02,lipniacki04,nelson04}.

We use the model introduced in \cite{lipniacki04} for the response
of $NF\kappa B$ module to a signal. This model is represented in
Fig.~\ref{lipni-model}. The first reaction of the model is the
activation of the kinase. In the absence of a signal the kinetic
constant of the activation reaction is zero $k_1 = 0$, meaning
that the kinase IKK remains inactive. The presence of a signal is
modeled by a non-zero activation constant $k_1
> 0 $, meaning that the kinase is activated.

We are interested in three characteristic times of the $NF\kappa
B$ model: the period and the damping time of the oscillations and
the largest relaxation time\footnote{The damping of the
oscillations is not necessarily the only relaxation process,
therefore the damping time is not necessarily equal to the largest
relaxation time.}.

We have studied numerically the dependence of these time scales on
the parameters of the model, which are the kinetic constants of the
reactions. The damping time $\tau_d$ and the largest relaxation time
$\tau_{max}$ were computed by linearizing the dynamical equations at
steady state. The period of the oscillation has a rigorous meaning
only for a limit cycle, when the oscillations are sustained. At a
Hopf bifurcation and close to it, the inverted imaginary part of the
conjugated eigenvalues crossing the imaginary axis provide good
estimate for the period. Another method for computing the period is
the direct determination of the timing between successive peaks. We
have noticed that in logarithmic scale, the differences between the
periods computed by the two methods were small, therefore we have
decided to use the first method, which is more rapid. A criterion
for the existence (observability) of the oscillations is the damping
time to period ratio. This ratio is infinite for self-sustained
oscillations, big for observable oscillations (when at least two
peaks are visible). A low ratio means over-damped oscillations. We
call the period an {\em observable} one, if the above ratio is
larger than one.

\subsection*{ $r$-robustness of the period}

\begin{figure}[t]
\centerline{ \hskip-0.5truecm
\includegraphics[width=14cm]{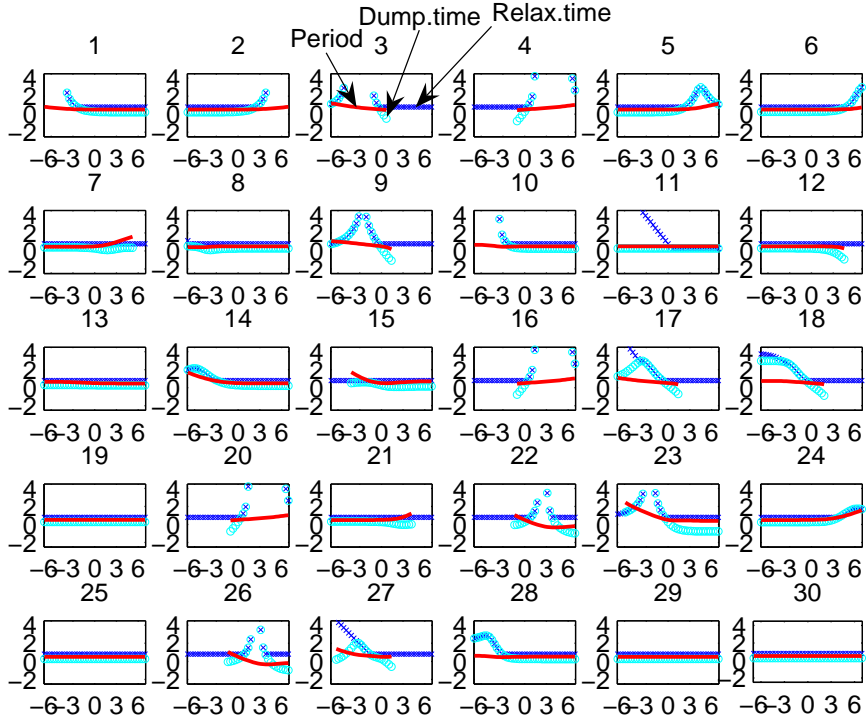}
}

\caption{ Log-log dependence of the characteristic times (blue
crosses: largest relaxation time, cyan circles: dumping time, red
line: period) on the scale factor that multiplies the value of one
parameter, while all the other parameters are fixed. The scale
factor varies from 0.001 to 1000 (from -6.9 to 6.9 in logarithm).
Oscillations have limited existence regions (outside these regions
they are overdamped; our subjective criterion for overdamping is a
damping time over period ratio smaller than 1.75). There are also
regions where oscillations are self-sustained. The limits of these
regions are Hopf bifurcation points, where the damping time and
the largest relaxation time diverge. Inside these regions, the
damping time and the largest relaxation time are infinite and not
represented.
 } \label{nfkb-temps}
\end{figure}

First, we have tested the $1$-robustness of the characteristic
times. Each parameter has been multiplied  by a variable, positive
scale factor (changing from 0.001 to 1000), all the other
parameters being kept fixed. The result can be seen in
Fig.~\ref{nfkb-temps}.

Large plateaus over which characteristic times are practically
constant correspond to robustness. The period of the oscillations is
particularly robust. For the damping time and the largest relaxation
time we have domains of substantial variation. There are two types
of such domains:

a) domains where $\D{\log(\tau)}{\log (k) } \approx -1 $.

b) domains where $|\D{\log(\tau)}{\log (k) } | > 1 $.

where $k$ is the variable parameter.

The first type of behavior is the same as the one of linear
networks. When changing only one parameter, there are domains in $k$
on which $k_{(r)}$ changes proportionally to $k$ (this corresponds
to $k=k_{(r)}$). In other words, this happens when the perturbation
acts on the ergodicity boundary. Outside these domains, $k_{(r)}$ is
constant, which means a plateau in the graph.

The second type of behavior exists only for non-linear networks and
is related to bifurcations. The variation of one parameter can bring
the system close to a bifurcation (for the $NF\kappa B$ model, this
is a Hopf  bifurcation) where the relaxation time diverges.

The in silico experiment shows that the largest relaxation time is
not $1$-robust; this time can be changed by modifying only one
parameter. The damping time has similar behavior being even less
robust (some plateaus of the largest relaxation time are higher
than the damping time, which continues to decrease; consider for
instance the effect of $k_9$ in Fig.~\ref{nfkb-temps}).

As also noticed by the biologists \cite{nelson04}, the period of the
oscillations is $1$-robust. We do not have a rigorous explanation of
this property. An heuristic explanation is the following. Close to
the Hopf bifurcation two conjugated eigenvalues $\lambda \pm i \mu $
of the Jacobian cross the imaginary axis of the complex plane;
$\lambda$ vanishes which explains the divergence of the relaxation
time, while $\mu$, whose inverse is the period, does not change
much. However, this is not a full explanation because it does not
say what happens far from the Hopf bifurcation point.

\begin{figure}[t]

\centerline{\includegraphics[width=12cm]{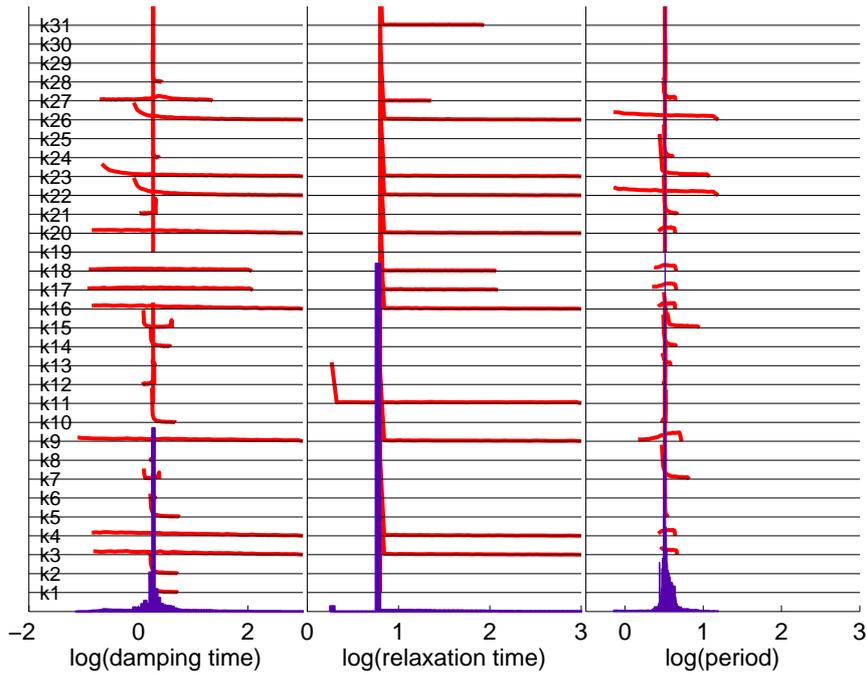} }

\caption{ Parameter sensitivity study; distributions of the
characteristic times when different parameters are, in turn,
multiplied by a log-uniform (between 0.1 and 10), random scale
factor, while all the other parameters are fixed. Distributions
corresponding to various parameters are spread out vertically. The
lower-most, bar-plotted distribution is the average of all the
distributions and corresponds to choosing randomly the parameter
to be modified. 1-robustness means that all distributions are
concentrated (their spread in log-scale is small). Weak
1-robustness means that only the average distribution is
concentrated.
 } \label{1-robustness}
\end{figure}

\subsection*{Parameter sensitivity}

Not all the parameters have the same influence on the
characteristic times. This can already be seen in
Fig.~\ref{nfkb-temps}. In order to quantify these differences we
have computed the distributions of the characteristic times when
one parameter is multiplied by a log-uniform random scale, all the
other parameters being fixed. This computation, whose results are
represented in Fig.~\ref{1-robustness} is also a first step
towards testing weak $r$-robustness.

Although rather robust, the period is not constant. Several
parameters induce relatively significant changes of this quantity.
In the order of increasing strength of their effect on the period,
these parameters are : $k_7, k_9, k_{15}, k_{23}, k_{22}, k_{26}$.
Among these, $k_{22},k_{26},k_9$ expressing the transcription
rates of mRNA-I$\kappa$B, the translation rates of I$\kappa$B, and
the biding rate of the kinase to the NF$\kappa$B-I$\kappa$B
complex are particularly interesting because by changing them, one
can increase and also decrease the period. These results confirm
and complete the findings of \cite{ihekwaba04}. The parameters
that have the greatest influence on the period are the kinetic
constants of the production module of I$\kappa$B :
$k_{22},k_{26}$. We should emphasize the very different timescales
of these reaction (one rapid $k_{22}=5\,\, 10^{-1}$ and the other
very slow $k_{26}=5\,\, 10^{-7}$). The strong influence of
NF$\kappa$B influx $k_{15}$ on the period, missed in
\cite{ihekwaba04}, is present here. Interestingly, the delay
produced in the transcription/translation module of A20 have
smaller effect on the period than the delay produced by the
I$\kappa$B production module. Less obvious is the effect of
$k_7,k_9$ (binding of IKK to I$\kappa$B or to the complex) on the
period, detected as important both here and in \cite{ihekwaba04}.

The damping time to period ratio represents a criterion for
observability of the oscillations. In order to increase the number
of visible peaks, one should increase the above ratio. Because the
period is  robust, this is equivalent  to increasing the damping
time. Figs.\ref{nfkb-temps},\ref{1-robustness} show that this is
possible in many ways by changing only one parameter (decrease of
$k_3$, or $k_9$, or $k_{17}$, or $k_{18}$, or $k_{23}$, or $k_{27}$
or increase of $k_4$, or $k_{16}$, or $k_{20}$, or $k_{22}$, or
$k_{26}$).

\begin{figure}[t]
\centerline{
\includegraphics[width=12cm]{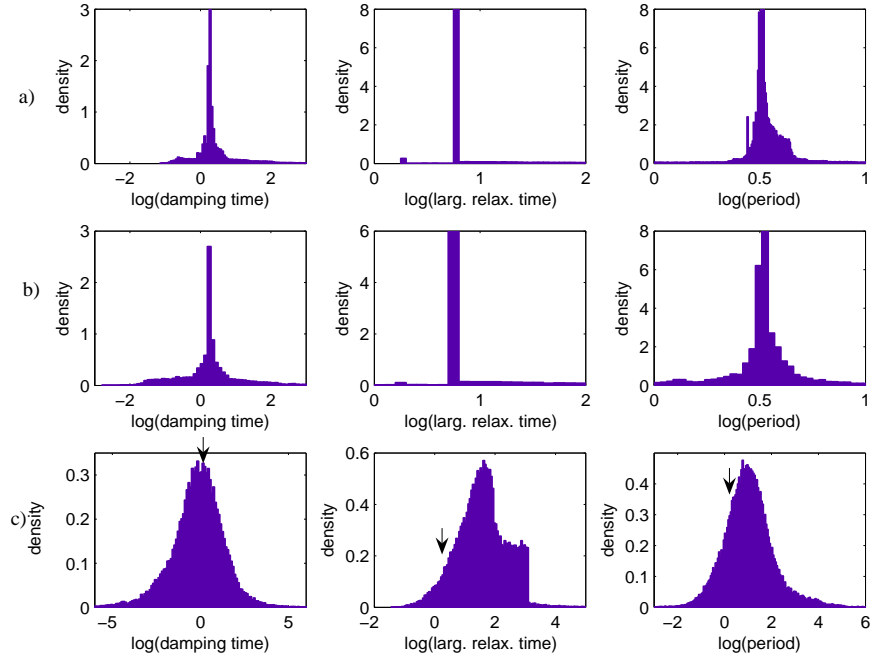}
}

\caption{ Distributions of characteristic times for log-uniform
(between 0.1 and 10), independent random scales multiplying the
kinetic parameters. We have changed a) one parameter, randomly
chosen; b) two parameters, randomly chosen; c) all the parameters.
The unperturbed values of the characteristic times are indicated
with arrows. The concentration of the distributions at a) and b)
shows that The period and the relaxation time are weakly 1- and
2-robust. The variation of all parameters produce long tailed
distributions (that can be fitted by log-generalized logistic
distributions) of the period and of the damping time, slightly
biased relative to the unperturbed values (the bias of the period
is positive, suggesting that it is easier to increase, than to
decrease the period by random perturbations). The distribution of
the relaxation time can be described as a mixture of a
log-generalized logistic, and of a log-beta distribution. Let us
remind that order statistics for log-uniform, independent
variables follow log-beta distributions.
 } \label{robustness}
\end{figure}

\subsection*{Weak $r$-robustness of all the characteristic times}

The divergence of the relaxation time close to a bifurcation does
not necessarily imply the absence of weak $r$-robustness or of
distributed robustness. The set of bifurcation points forms a
manifold in the space of parameters, of codimension equal to the
codimension of the bifurcation; in general, this set has zero
measure ~\footnote{Stochastic cellular automata provide an
interesting counter-example : the NEC automaton of Andrei Toom
\cite{toom}}. The probability of being by chance close to a
bifurcation is generally small.

We have tested the weak $r$-robustness of the characteristic
times, by using independent, log-uniform distributions of the
parameters over 2 decades interval. All the three relaxation times
are weakly $r$-robust when $r$ is small (see
Fig.~\ref{1-robustness}, Fig.~\ref{robustness}a,b). Thus, although
controlable (there are critical parameters), the system is somehow
robust. Only an informed choice of the right targets has an
effect, random choice of a small number of targets is inefficient.

We have also tested distributed robustness. Interestingly, when
all the parameters take independent log-uniform values, the
distributions of characteristic times, are much broader than the
ones induced by changing a small number of parameters (see
Fig.~\ref{robustness}c). Neither the longest relaxation time, nor
the damping time have distributed robustness. Nevertheless, the
period is slightly more robust that the other characteristic
times. In logarithmic scale, the distributions of the dumping time
and of the period have  exponential tails, with different decays
rates towards $\infty$ and $-\infty$. These distributions  (a
possible fit is by generalized log-logistic distributions) have
longer tails (in log scale) than log-normal distributions that are
sometimes observed in biology
\cite{random-periods,konishi,bengtsson,furusawa,lognormal}. The
tails are also longer than the ones of the Tracy-Widom
distribution characterizing largest eigenvalues of certain classes
of random matrices \cite{Tracy96,Soshnikov02}. These long tails
are related to the critical retardation phenomena
\cite{GorSloRelax} close to the Hopf bifurcation (see also
Fig.~\ref{nfkb-temps}). The distribution of the relaxation time
can be seen as a mixture between a log-normal and a
log-generalized logistic distribution.

\begin{figure}[H]

\centerline{
\includegraphics[width=8cm]{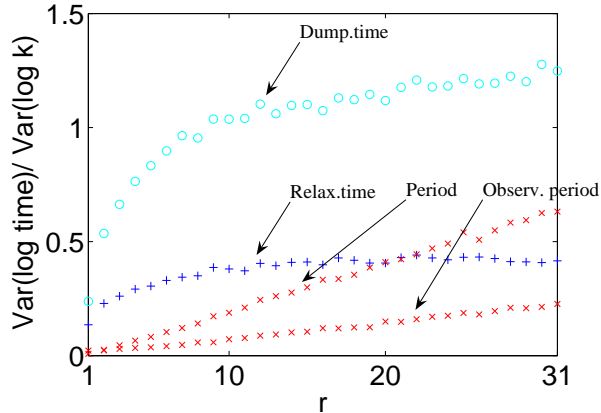}
}

\caption{ Log variance of the characteristic times as function of
$r$, the number of perturbed parameters. The choice of the $r$
parameters is random (uniform) and the values of the random scales
are independent, log-uniform (between 0.1 and 10). Some statistical
samples correspond to overdamped oscillations ( dumping time /
period ratio $<$ 1); these samples were rejected when computing the
log-variance of the observable period.
 } \label{robustness_index}
\end{figure}

In order to quantify $r$-robustness we have plotted in
Fig.~\ref{robustness_index} the dependence of the log variance of
the characteristic times on $r$, the number of the perturbed
parameters. For a property governed by simplex concentration (for
instance the largest relaxation time of a linear network) one
would expect the existence of a well defined robustness index. For
$r$ values smaller than or equal to the robustness index, the
system is $r$-robust, while for larger values it is not. This
would imply a large increase of the log-variance with $r$ before
the robustness index, and a much slower increase after. In
Fig.~\ref{robustness_index} this is satisfied by the dumping time
and by the largest relaxation time (robustness index between 1 and
7). The period has a completely different behavior, its
log-variance being proportional to  $r$. This means that no
r-value is privileged, and in order to keep the period robust one
should avoid multiple perturbations. This is more likely a cube
concentration phenomenon, when the effect of different parameters
is cumulative.

However, the log-variance of the period can not increase
indefinitely. Even when perturbing all the parameters of the system,
this log-variance remains about half smaller than the log-variance
of the dumping time; it is even smaller if we consider only
observable oscillations (big dumping time to period ratio).
Concentration means that the slope of the linear dependence of the
log-variance of the period on $r$
 will tend to zero in an hierarchy of
models of increasing complexity. The limit behavior can not be
tested here because we analyze only one model, and not an hierarchy.

\section{Discussion and conclusions}

We demonstrated the possibility of a new kind of robustness of
biological systems. This type of robustness has geometrical
origin, being related to the high dimension in which variability
sources act. There are two basic types of such geometrical
effects: cube type of concentration, or simplex type.

The classical example of the cube concentration gives the central
limit theorem, when the robust property is the sum of many ($n$),
independent contributions. For concentration of this type, the
relative standard deviation decreases as $1/\sqrt{n}$. The
classical example of the simplex concentration, is the situation
when the robust property depends on the $k$-st in order effect
(parameter) in a collection of of many ($n$) effects (parameters),
for example, on the smallest, the largest or the second one. For
concentration of this type, the relative standard deviation
decreases much faster, as $1/n$. We have provided strong arguments
that the robustness of the largest relaxation time in large
multiscale networks with hierarchically distributed timescales is
of the simplex type. The concentration of the period of a
nonlinear oscillator seems to be of the cube type. This is
coherent with the fact that the period results from the cumulative
effect of various delay sources.

We have also defined the concepts of distributed robustness and
$r$-robustness that occur naturally in cell physiology and molecular
biology.  The robustness index is the maximal integer $r$ such that
the system is $r$-robust. For analysis of networks we discussed two
types of noise: random noise in constants and destruction of links.
The necessity of robustness to both types leads to a new
combinatorial problem: How to create a minimal network that has
sufficiently long routes (the length of each route is $>L$), and, at
the same time, sufficiently many routes:   after destruction of
$D_{\rm l}$ links and $D_{\rm n}$ nodes there remains at least one
route in the network.

In a recent work Rand \cite{Rand06} introduces the flexibility
dimension that quantifies the range of evolution of clocks. This
notion applies to multitask evolution, simultaneously fulfilling
several objectives. By using linear response theory the authors
proposed a method to compute the directions in the characteristic
space that are not robust to changes of the parameters: the
flexibility dimension is the largest linear space of characteristics
that contains non-robust directions. Our notion of robustness index
is different because it does not follow from linear response and
more importantly it applies to parameters and not to characteristics
(it is the maximum number of parameters that can be changed without
sensible effect on the characteristics). Nevertheless it seems that
the flexibility dimension and the robustness index have properties
in common: they are both small for simple networks and they tend to
be increased by the loop complexity and by the unevenness of the
lifetimes of various species.

Concerning the analyzed example, several conclusions are important.
NF$\kappa$B dynamics belongs to the category of ultradian
oscillators. As for circadian oscillators \cite{Rand06}, the period
of the oscillations is a relatively robust property. Even if the
biological role of these oscillations has not yet been proven (for
some conjectures the reader can refer to \cite{nelson04}), it is
important to known that the robustness applies to different
timescales. A specificity of the the NF$\kappa$B system is the
proximity to a bifurcation. Two  non-linear phenomena could be
relevant for the behavior of the signalling system : the critical
retardation and the excitability. The first property would produce
long tail distributions of the damping time of the oscillations and
interesting possibilities for synchronization. The second property
could raise the efficiency of the regulatory role of NF$\kappa$B by
increasing the amplitude of its response to signals.

The robustness of a system is related to its complexity. More
precisely, it is given by two levels of complexity: the one on
which the sources of variability act and the one of its dynamics.
The higher is the level of complexity where the variability acts
and the lower is the level of dynamical complexity, the higher is
the robustness. In order to test the concentration from high
dimension rigorously one needs to build an hierarchy of models
obtained one from another by model reduction. Parameters of
simpler models in the hierarchy are functions of packages of
parameters (``atoms") of more complex models. Independent
perturbations of the atoms  produce less variability than overall
perturbation of packages. These ideas will be presented in detail
in a future work.

{\small
\bibliography{complex,markov-bib,modules}
\bibliographystyle{alpha}
}

\end{document}